\documentclass[twocolumn,secnumarabic,amssymb, nobibnotes, superscriptaddress,aps, prd]{revtex4-1}

\newcommand{\bea}{\begin{eqnarray}}
\newcommand{\ena}{\end{eqnarray}}
\newcommand{\be}{\begin{equation}}
\newcommand{\ee}{\end{equation}}
\newcommand{\beann}{\begin{eqnarray*}}
\newcommand{\enann}{\end{eqnarray*}}

\usepackage{eurosym}
\usepackage{graphicx}
\usepackage{amsmath}
\usepackage{mathrsfs}
\usepackage{bm}
\usepackage{color}
\usepackage{amsmath}
\usepackage{amssymb}

\usepackage{multirow}
\usepackage{array}
\usepackage{subfig}

\usepackage[utf8]{inputenc}
\usepackage{graphics,graphicx}
\usepackage{amsmath,amssymb,amsfonts,latexsym,cancel}
\usepackage{amsmath,amssymb}

\begin{document}

\title{Normal and Quasinormal Modes of Holographic Multiquark Star}

\author{Supakchai Ponglertsakul}
\email{supakchai.p@gmail.com}
\affiliation{Strong Gravity Group, Department of Physics, Faculty of Science, Silpakorn University, Nakhon Pathom 73000,
	Thailand}

\author{Piyabut Burikham}
\email{piyabut@gmail.com}
\affiliation{High Energy Physics Theory Group, Department of Physics, Faculty of Science, Chulalongkorn University, Bangkok 10330, Thailand}

\author{Sitthichai Pinkanjanarod}
\email{quazact@gmail.com}
\affiliation{ Department of Physics, Faculty of Science, Kasetsart University, Bangkok 10900, Thailand}

\date{\today}

\begin{abstract}

The quadrupole normal-mode oscillation frequency $f_{n}$ of multiquark star are computed for $n=1-5$.  At the transition from low to high density multiquark in the core region, the first 2 modes jump to larger values, a distinctive signature of the presence of the high-density core.  When the star oscillation couples with spacetime, gravitational waves~(GW) will be generated and the star will undergo damped oscillation. The quasinormal modes~(QNMs) of the oscillation are computed using two methods, direct scan and WKB, for QNMs with small and large imaginary parts respectively. The small imaginary QNMs have frequencies $1.5-2.6$ kHz and damping times $0.19-1.7$ secs for multiquark star with mass $M=0.6-2.1 M_{\odot}$~(solar mass). The WKB QNMs with large imaginary parts have frequencies $5.98-9.81$ kHz and damping times $0.13-0.46$ ms for $M\simeq 0.3-2.1 M_{\odot}$. They are found to be the fluid $f-$modes and spacetime curvature $w-$modes respectively. 

\end{abstract}
\maketitle

\section{Introduction}\label{sec-introd}

The late-time gravitational waves signal from compact sources is prominently dominated by characteristic ringdown phase. This phase is described by a so-called quasinormal modes (QNMs). In principle, one can determine the nature of the source by measuring damping time of gravitational waves amplitude. Moreover, linear stabilities of compact objects can also be verified by the QNMs. An exponentially decay of perturbation mode indicates that a perturbed object is stable under a linear perturbation. The studies of black hole quasinormal modes can be traced back to 1970, where Vishveshwara calculates an oscillation of a gaussian wavepacket around the Schwarzschild black hole \cite{Vishveshwara:1970zz}. It turns out that, the frequency and damping of these oscillations are solely characterized by its mass. Since then, numerous of a similar study have been explored with various types of black holes and fields (see \cite{Konoplya:2011qq, Kokkotas:1999bd, Ferrari:2007dd} for a nice review on this subject).  Beyond general relativity, QNMs of black holes in modified gravity are studied in great numbers. For instance, extended analyses for black hole/string in massive gravity and generalized spherical symmetric background~\cite{Burikham:2020dfi,Wuthicharn:2019olp,Ponglertsakul:2018rot,Ponglertsakul:2018smo,Burikham:2017gdm} demonstrate rich structure of the black hole QNMs and remarkable connections to the Strong Cosmic Censorship.

Similarly, the study of stellar pulsation in general relativity has a long history \cite{ Thorne1967, Thorne:1968zz, PriceThorne, Thorne, Chanmugam1977, Ipser1979, Lindblom:1983ps, Kojima1991}. In Ref.~\cite{ Kokkotas:1992xak, Andersson:1995wu}, a new family of oscillation modes of neutron stars are discovered i.e. $w-$modes or spacetime modes. The $w-$modes are much closer to black hole QNMs than the fluid modes, i.e., $f, p, g$ and $r-$modes. In addition, the $w-$modes are oscillating with very rapid damping time. We refer an interested reader to Ref.~\cite{Kokkotas:1999bd} for a classification of neutron stars oscillation modes including sub-family of the spacetime modes. Moreover, the investigation of gravitational waves signal emitted from neutron stars can prove to be useful. It is expected that gravitational radiations from neutron stars carry information about the star’s internal structure such as its radius, density and properties of nuclear and quark matter~\cite{Tsui:2004qd,Benhar:1998au, Kokkotas:1999mn}. 

There have been a series of investigation on the physical properties of massive neutron star~(NS) with exotic quark-matter core in the form of multiquark~(MQ) phase from the holographic Sakai-Sugimoto~(SS) model~\cite{bhp,bch,Pinkanjanarod:2020mgi,Pinkanjanarod:2021qto,Burikham:2021xpn}.  There are also other types of holographic models, e.g. top-down models D3-D7, D4/D8/$\bar{\rm D8}$ and a bottom-up V-QCD model considered in the study of neutron stars and hybrid stars~\cite{Hoyos:2021uff,Jarvinen:2021jbd,Hoyos:2016zke,Jokela:2021vwy,Alho:2020gwl}. In the top-down SS model, the dual gauge theory has many similar properties to the QCD, e.g. confinement/deconfinement phase transition, chiral symmetry breaking/restoration. Baryon can be added to the model by introduction of baryon vertex attached to $N_{c}$ strings from the AdS boundary. Multiquark states can be constructed by adding equal numbers of string in and out of the baryon vertex~\cite{bch}. It is interesting that when the bulk spacetime contains horizon and the chiral symmetry is broken by the connecting D8-$\overline{\rm D8}$ configuration, the multiquark phase is found to be the most thermodynamically preferred nuclear phase in the SS model for moderate temperature~(less than trillion Kelvins) and high density~\cite{bch}. Such physical condition is expected to be found in the core of massive compact astrophysical object such as neutron star. Consequently, aged massive NS above 2 solar masses is likely to contain the multiquark core and nuclear crust according to the SS model. 

At the beginning of neutron star/hybrid star formation, e.g. after supernovae explosion, white dwarfs merging, excess mass intake of a neutron star; the resulting ultradense compact object could have an extremely high temperature, comparable to deconfinement phase transition temperature. In such circumstances, the compact star resulting from the collapse could be entirely in the deconfined MQ phase~(or even contain quark-gluon plasma~(QGP) core if core temperature is sufficiently large), where the coupling between quarks is still sufficiently strong to form bound states even when the quarks are deconfined within the much larger deconfined region. It is thus interesting to investigate the physical properties of massive compact stars entirely consisting of the MQ matter by assuming temperature at the star surface to be higher than the transition temperature between multiquark and ordinary nuclear matter. In Ref.~\cite{Pinkanjanarod:2021qto}, radial pulsations of the holographic MQ star have been analyzed and six characteristic frequency modes are determined. 

In this work, we calculate the normal and quasinormal modes~(QNMs) of holographic MQ stars in the SS model, in both the fluid-oscillation~(Newtonian $f-$modes) and spacetime~($w-$modes) modes. By comparison to observations, these modes can be used to identify the MQ star/core with holographic equations of state in addition to other specific physical properties studied in Ref.~\cite{Pinkanjanarod:2020mgi,Pinkanjanarod:2021qto,Burikham:2021xpn}. The results can also be compared with frequencies of other possibilities of massive compact object such as strangeon stars~(SS)~\cite{Li:2022qql} or other phases with different equations of state such as color-superconductivity~(CSC)~\cite{Fukushima:2007fc,Feng:2009vt}~(see Ref.~\cite{Alford:2007xm} and references therein).   

The work is organized as the following. Section \ref{sec-th} presents the equations of state~(EoS) for holographic multiquark matter in the SS model and the perturbation equations of motion.  Section \ref{SectNM} considers normal modes of fluid MQ star oscillation.  QNMs for fluid and spacetime modes are calculated in Section \ref{SectQNM}. Section \ref{sec-con} concludes our work. 

\section{Theoretical Setup: Equations of state and Equation of motion}  \label{sec-th}

\subsection{Equations of State}  \label{sec-eos}

``Neutron stars" generically cool down rather quickly to temperatures lower than $0.1$ MeV~($\sim 10^9$ K). At such moderate temperatures below the quark-gluon plasma formation, the coupling of strong interaction could still be strong. With extreme densities inside the neutron star, hadronic matters, e.g. neutrons and protons could not withstand extreme pressure, and the boundary between hadrons could overlap. Quarks inside one hadron might leak into others and could form multiquark bound states within an even larger confinement radius inside the star. During the earlier stage of the cool down process, the entire untra-dense star could have temperature in the range $10^{9}~{\rm K}<T<10^{12}$ K, where the holographic MQ phase is the most thermodynamically preferred phase~(see Ref.~\cite{bch,Pinkanjanarod:2020mgi}). During this period, the young ``neutron star" could actually be the MQ star. And in the later time, the aged ``NS" could actually be the hybrid star with MQ core. Another scenario where MQ star could be formed is when the massive NS gains more mass, collapses and heats up until most of the star is in the MQ phase.  

At present, there is still no effective theory derived from QCD that could describe the hydrodynamic and thermodynamic behaviour of the MQ states due to their strongly coupled nature. We thus adopt the EoS of MQ matter from holographic SS model originally computed in Ref.~\cite{bhp}. At high densities, the holographic MQ phase was found to be more thermodynamically preferred over the stiff nuclear matter described by chiral effective theory or CET in Ref.~\cite{Pinkanjanarod:2020mgi} and the KSZ models~\cite{Kim:2007zm,Kim:2007vd} in Ref.~\cite{Burikham:2021xpn}. It should be noted that the MQ phase can be extended to include CSC when diquark condensate is formed at relatively low temperature, and include strange quark flavor to address the possibility of strange matter emergence, with  and without the presence of strong coupling effects.

Correspondingly, the multiquark matter inside the star can be described by relations between pressure $P$ and mass density $\rho$, written as a function of the number density $n$, as described in Ref.~\cite{bhp}. The EoS for the small $n$~(``mql'') multiquark can be expressed in the dimensionless form as 
\bea
P &=& a n^{2}+b n^{4},  \notag \\
\rho  &=& \mu_{0}n+a n^{2}+\frac{b}{3}n^{4},  \label{eosmq1}
\ena
where $\mu_0$ is the initial value of the chemical potential for the multiquark phase, while $a, b$ are the parameters associated with the small $n$ holographic multiquark EoS. On the other hand, for large $n$~(``mqh'')
\bea
P &=& k n^{7/5},  \notag \\
\rho  &=& \rho_{c} +\frac{5}{2}P+\mu_{c}\left(n-n_{c}\right) \notag \\
&&+kn_{c}^{7/5}-\frac{7k}{2}n_{c}^{2/5}n,  \label{eosmq2}
\ena
where a critical mass density, chemical potential, and number density at the transition from large to small multiquark number density, represented by $\rho_c$, $\mu_c$, and $n_c$, respectively.
\begin{table}[ht]
	\centering	
	\begin{tabular}{ |c|c|c|c|c|c|c|c| }
		
		\hline
		&&&&&&&\\
		
		$n_s$ & $a$ & $b$ & $\mu_0$ & $k$ & $\rho_c$ & $\mu_c$ & $n_c$ \\
		
		&&&&&&&\\
		
		\hline
		
		&&&&&&&\\
		0 & 1 & 0 & 0.17495 & $10^{-0.4}$ & 0.0841077 & 0.564374 & 0.215443   \\
		&&&&&&&\\
		
		\hline
		
		&&&&&&&\\
		0.3 & 0.375 & 180.0 & 0.32767 & $10^{-0.4}$ & 0.0345996 & 0.490069 & 0.086666 \\
		&&&&&&&\\
		
		\hline
		
	\end{tabular}
	
	\caption{Parameters associated with EoS expressed in Eqs.(\ref{eosmq1}) and (\ref{eosmq2}) in dimensionless units}
	
	\label{tabI}
	
\end{table}

Eqs.(\ref{eosmq1}) and (\ref{eosmq2}) are expressed in dimensionless form where all parameters are provided in Table \ref{tabI}. For both $n_{s} =0$ and $0.3$ cases, the parameter $k=10^{-0.4}$. This implies that the MQ at high density is independent of the colour charges as they have similar characteristics. Notably, these EoS are shown to be insensitive to temperature in the range $10^{9}~{\rm K}<T<10^{12}$ K~\cite{bhp}. The EoS for multiquark depends on two free parameters: the colour charge of the multiquark state $n_{s}$ and the energy density scale $\epsilon_{s}$~\cite{bch, bhp}. Converting to SI units, the pressure $P$ and mass density $\rho$ are proportional to the energy density scale $\epsilon_{s}$. As a result, the pure MQ-star mass and radius have the same scaling $M, R \sim \epsilon_{s}^{-1/2}$. Remarkably, the compactness $M/R$ is thus unaffected by $\epsilon_{s}$. In this work, we will set $\epsilon_{s}=26$ GeV fm$^{-3}$.

\subsection{Equations of motion}  \label{SectEOM}

We use the convention of Ref.~\cite{Kokkotas:1992ka, Andersson:1995wu}.  The metric is expressed as
\bea
ds^{2}&=&-e^{\nu}(1+r^{\ell}H_{0}Y^{\ell}_{m}e^{i\omega t})dt^{2}-2i\omega r^{\ell+1}H_{1}Y^{\ell}_{m}e^{i\omega t}dt~dr \notag \\
&&+ e^{\lambda}(1-r^{\ell}H_{0}Y^{\ell}_{m}e^{i\omega t})dr^{2} \notag \\
&&+r^{2}(1-r^{\ell}KY^{\ell}_{m}e^{i\omega t})(d\theta^{2}+\sin^{2}\theta~d\phi^{2}),
\ena
where $\nu,\lambda,H_0,H_1,K$ are functions of radial coordinate $r$ and the spherical harmonics is denoted by $Y^{\ell}_{m}$. The fluid 4-velocity sourcing the spacetime perturbation for even parity modes are given by~\cite{thcam}~(with $r^{\ell}$ rescaling),
\bea
u^{0}=e^{-\nu/2}(1-\frac{1}{2}r^{\ell}H_{0}Y^{\ell}_{m}e^{i\omega t}),&&~u^{r}=r^{\ell -1}e^{-(\nu +\lambda)/2}\partial_{t}W~Y^{\ell}_{m}, \notag \\
u^{\theta}=-r^{\ell-2}e^{-\nu/2}\partial_{t}V\partial_{\theta}Y^{\ell}_{m},&&~u^{\phi}=0,
\ena
where $W = W(t,r)$ and $V = V(t,r)$. Generally, perturbed Einstein field equations yield five first order differential equations for $H_0,H_1,K,W,V$. However, they are not all linear independent. In fact, the  Einstein equation implies the following \cite{Kokkotas:1992xak}
\begin{align}
\left(2M+Nr+\bar{Q}\right)H_0 &= - \Bigg[ \left( N+1\right)\bar{Q}-\omega^2 r^3 e^{-\left(\lambda+\nu\right)} \Bigg]H_1 \nonumber \\ 
&~~~+ \Bigg[ Nr - \omega^2 r^3 e^{-\nu} - \frac{e^{\lambda}}{r}\bar{Q}\left(2M \right. \nonumber \\
&~~~\left. -r+\bar{Q}\right) \Bigg]K + 8\pi r^3 e^{-\frac{\nu}{2}}X, \label{constrain}
\end{align}
with $\bar{Q}=M+4\pi r^3 P$ and $N=(\ell-1)(\ell+2)/2$. The new function $X$ is introduced \cite{Lindblom:1983ps}
\bea
X&\equiv&\omega^{2}(P+\rho)e^{-\nu/2}V+\frac{(P+\rho)}{2}e^{\nu/2}H_{0}-\frac{P'}{r}e^{(\nu-\lambda)/2}W. \notag \\ \label{defX} 
\ena
The equations of motion governing the perturbations of fluid in a spherically symmetric star with no rotation are given by the Einstein field equations,
\bea
H'_{1}&=&\frac{e^{\lambda}}{r}\Bigg( -[(\ell+1)e^{-\lambda}+2\frac{M}{r}+4\pi r^{2}(P-\rho)]H_{1} +H_{0}+K \notag \\
&&-16\pi(P+\rho)V \Bigg),  \notag  \\
K'&=&\frac{1}{r}\Bigg( H_{0}+(N+1)H_{1}-(\ell+1-\frac{r\nu'}{2})K  \notag \\
&& -8\pi(P+\rho)e^{\lambda/2}W \Bigg), \notag  \\
W'&=&-(\ell+1)Wr^{-1}+re^{\lambda/2}\Bigg( \frac{1}{2}H_{0}+K+(\gamma P)^{-1}e^{-\nu/2}X  \notag \\
&&-\ell(\ell+1)r^{-2}V\Bigg),  \notag  \\  
X' &=& -\frac{\ell}{r} X + \frac{1}{2r}\left(\rho+P\right)e^{\frac{\nu}{2}}\Bigg[ \left(1-\frac{\nu'r}{2}\right)H_0 + \bigg(r^2\omega^2 e^{-\nu} \notag \\
&& + \left(N+1\right)\bigg)H_1 + \left(\frac{3}{2}\nu'r-1\right)K - \frac{\ell(\ell+1)\nu'}{r}V \notag \\
&& - \bigg( 8\pi \left(\rho+P\right)e^{\frac{\lambda}{2}}+2\omega^2e^{\frac{\lambda}{2}-\nu} - r^2\left(\frac{\nu'}{r^2e^{\frac{\lambda}{2}}}\right)' \bigg)W \Bigg]. \notag \\   \label{eom}
\ena
Remark that, we can eliminate $H_0$ and $V$ using \eqref{constrain} and \eqref{defX} respectively from the perturbation equations \eqref{eom}. Thus we obtain four first order differential equations for $\{H_1,K,W,X\}$.

We will solve these perturbation equations with appropriated boundary conditions. As a result, one obtains specific frequency $\omega$. For the calculation of normal modes, the boundary condition at the star surface $r=R$ is simply $X(R)=0$~(zero pressure and density at the surface) and there is no need to do the matching with the outer region of the star. The normal modes correspond to real frequency. Boundary conditions and numerical procedure for obtaining QNMs and their corresponding quasinormal frequency can be found below in Sect.~\ref{SectQNM}.

\section{Normal modes of Multiquark Star}   \label{SectNM}

\begin{figure}[h]
	\centering
	\includegraphics[width=0.50\textwidth]{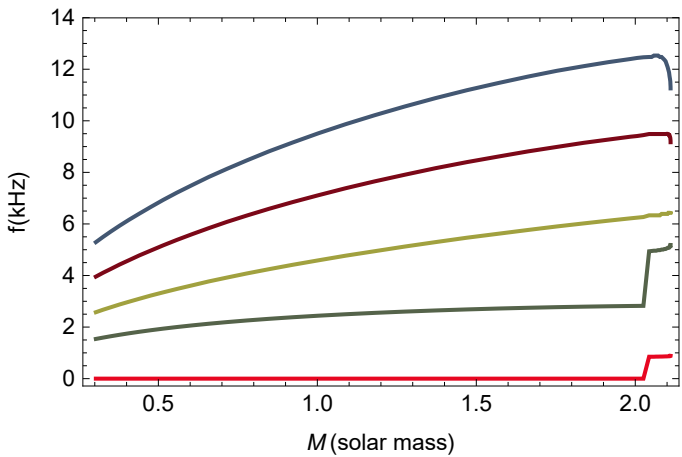}
	\caption{Quadrupole oscillation frequency $f_{n}, n=1,2,3,4,5$ vs. $M$ of the MQ star.}
	\label{fMfig}
\end{figure}

\begin{figure}[h]
	\centering
	\includegraphics[width=0.5\textwidth]{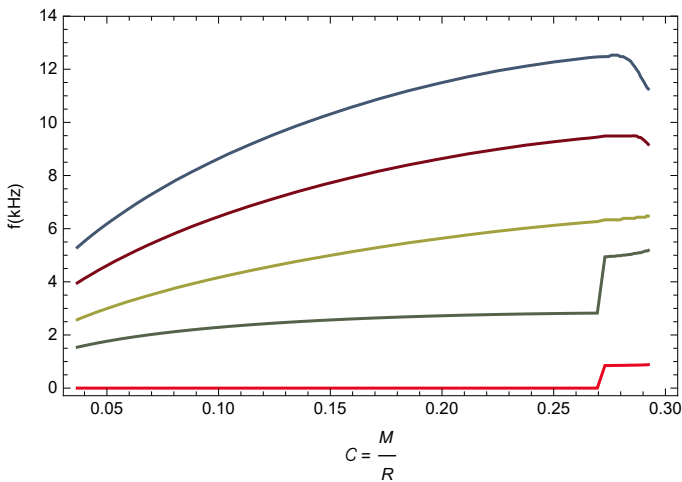}
	\caption{Quadrupole oscillation frequency $f_{n}, n=1,2,3,4,5$ vs. compactness $C$ of the MQ star.}
	\label{fCfig}
\end{figure}

In this section, the normal modes quadrupole oscillations and QNMs of the multiquark star are numerically calculated.  The normal mode quadrupole oscillations are computed under the assumption that the energy loss to the gravitational waves is negligible and the frequency $f=\omega/2\pi$ is purely real.  These Newtonian modes are simply fluid nonradial oscillations confined by gravity of the star with boundary conditions $X=0, r\geq R$. In Fig.~\ref{fMfig} and Fig.~\ref{fCfig}, the 5 lowest frequency modes ($n=1-5$) are displayed as a function of MQ star’s mass and star’s compactness $C\equiv M/R$ respectively~(note that the MQ star with mass smaller than $1.4 M_{\odot}$ is most likely hypothetical but we choose to present them for comparison to the typical NS with other nuclear EoS). The frequencies are found to be monotonically increasing with $M$ and $C$. At high mass $M \gtrsim 2 M_{\odot}$ and compactness when the star has the high density ``mqh" core, the behavior of the frequencies become non-trivial. The high-density core acquires its own fundamental oscillation resulting in the appearance of the lowest mode $f_{1}$.  At the same time, the second mode jumps to higher value while $f_{4}$ and $f_{5}$ start to decrease with increasing $M, C$ at higher masses while $f_{3}$ coincidentally stays monotonic. In the presence of very small Newtonian damping, these normal modes obtain very small imaginary part in the frequency and become the $f-$modes discussed in Section~\ref{SectQNM}.

In comparison, typical $f-$mode frequencies of the neutron star with nuclear matter content are in the range $1.5-3$ kHz~\cite{Kokkotas:1999bd,Lindblom:1983ps} while the $p_{1}-$mode frequencies are around $4-7$ kHz. For the multiquark star with masses around $1.4$ solar masses, the $f_{2}$ is the lowest mode and its value is $2.6-2.7$ kHz. The multiquark oscillation has relatively high fundamental frequency comparable to the nuclear matter with high mean density. This is consistent with the fact that the multiquark phase consists of extremely dense bound states of quarks. The ``mql" multiquark behaves very similar to baryon. Only when the density becomes even higher that ``mqh" multiquark EoS emerges and the most fundamental frequency $f_{1}$ appears in the massive multiquark star above $2$ solar masses.

\section{Quasinormal modes of Multiquark Star}    \label{SectQNM}

For compact objects with extreme density, quadrupole oscillations of star can couple to the spacetime generating GWs which carry energy away.  In this case, the oscillation will be damped resulting in the QNMs of the dense massive star.  Depending on the physical modes, coupling between spacetime and fluid content of the star could be drastically different.  While normal-mode quadrupole oscillations are $f-$modes, oscillatory modes with small spacetime-fluid coupling could have small imaginary parts of $\omega$.  

The QNMs are calculated using two methods, the direct scan and WKB. Both methods find solution with zero incoming waves. Direct scan is used to find the QNMs with very small ${\rm Im~\omega}$, the ratio of incoming to outgoing waves are in the order of $10^{-6}$ at far distance $r=55 ~{\rm Re}~\omega$. For WKB, the outer wave solutions of QNMs are verified that they contain less than $10^{-3}$ of the incoming/outgoing waves ratio at far distance $r>50 ~{\rm Re}~\omega$. Fig.~\ref{QNMfig1}, \ref{QNMfig2} and \ref{QCfig} show real and imaginary parts of $\omega$ in dimensionless unit~(note that the MQ star with mass smaller than $1.4 M_{\odot}$ is most likely hypothetical but we choose to present them for comparison to the typical NS with other nuclear EoS).  The value of ${\rm Re}~\omega$ can be translated to the frequency $f$ in the SI units by the conversion factor $f_{\rm con}\equiv 1.73603 \sqrt{\epsilon_{s}/({\rm GeV fm^{-3}})}=8.85206$ kHz for $\epsilon_{s}=26$ GeV$/{\rm fm}^3$.  The damping time $\tau\equiv 1/{\rm Im}~\omega$ can be translated to time unit by the conversion factor $t_{\rm con}\equiv 0.01797943211$ ms~(scales with $1/\sqrt{\epsilon_{s}}$).

\subsection{The inner solution of the star}  \label{SectInnerSol}

To ensure the condition $X(R)=0$ is satisfied, we use the same procedure as Ref.~\cite{Lindblom:1983ps} to solve Eqs.~(\ref{eom}) in the form
\be
\frac{d \mathbf{Y}}{dr} = \mathbf{P}(r,\ell,\omega)\cdot \mathbf{Y},
\ee
where $\mathbf{Y}=(H_{1}, K, W, X)$ and the matrix $\mathbf{P}$ can be read off from (\ref{eom}) after using constraints from the equations of motion to eliminate $V, H_{0}$. Then three independent solutions with $X(R)=0$ are numerically solved from the surface to the radius $r=R/2$ and two independent solutions are solved from the center out to $r=R/2$. The general solutions can be expressed as  
\bea
\mathbf{Y}_{\rm in}(R/2\leq r \leq R)&=&a_{1}\mathbf{Y}_{1}(r)+a_{2}\mathbf{Y}_{2}(r)+a_{3}\mathbf{Y}_{3}(r),  \notag \\
\mathbf{Y}_{\rm in}(0\leq r \leq R/2)&=&a_{4}\mathbf{Y}_{4}(r)+a_{5}\mathbf{Y}_{5}(r).
\ena  
The physical inner solution requires the matching
\bea
a_{1}\mathbf{Y}_{1}(R/2)&&+a_{2}\mathbf{Y}_{2}(R/2)+a_{3}\mathbf{Y}_{3}(R/2) \notag \\
&&= a_{4}\mathbf{Y}_{4}(R/2)+a_{5}\mathbf{Y}_{5}(R/2).
\ena

\subsection{The outer solutions}

At surface of the star, the inner and outer solutions are matched via the transformation related to Zerilli function $Z(r_{*})$
\be
\begin{pmatrix}
0 & 1\\
1 & 0 \\
\end{pmatrix}
\begin{pmatrix}
r^{-(\ell+1)}H_{1}(r)\\
r^{-\ell}K(r) \\
\end{pmatrix}
=
\begin{pmatrix}
f(r) & g(r)\\
h(r) & k(r) \\
\end{pmatrix}
\begin{pmatrix}
Z(r_{*})\\
Z'(r_{*}) \\
\end{pmatrix},
\ee
where $Z'(r_{*}) \equiv \displaystyle{\frac{dZ}{dr_{*}}}$,
\bea
f(r)&=&\frac{N(N+1)r^{2}+3NMr+6M^{2}}{r^{2}(Nr+3M)}, g(r)=1, \notag \\
h(r)&=&\frac{-Nr^{2}+3NMr+3M^{2}}{(r-2M)(Nr+3M)}, k(r)=\frac{-r^{2}}{r-2M}, \notag
\ena
and the tortoise coordinate $r_{*}$ is
\be
r_{*}=r+2M\log \Big(\frac{r}{2M}-1\Big). 
\ee
Outside the star, there is no source of matter and the spacetime resembles the Schwarzschild metric. Therefore the perturbation equations reduce to a single wave-like equation \cite{Zerilli:1970se}.  Generically, the wave equation~(Zerilli's equation) in the exterior of the star is given by 
\be
\frac{d^{2}Z}{dr_{*}^{2}}+(\omega^{2}-V(r))Z = 0,  
\ee
where the effective potential is
\bea
V(r) = &&\frac{2(1-2M/r)}{r^{2}(Nr+3M)^{2}}  \\
&&\times \Big( (N+1)N^{2}r^{3}+3N^{2}Mr^{2}+9NM^{2}r+9M^{3} \Big). \notag
\ena
For real $\omega$, there are two linearly independent solutions to the Zerilli's equation,
\bea
Z_{-}(r_{*}) &=& e^{-i\omega r_{*}}\sum_{j=0}^{\infty}\alpha_{j}r^{-j},  \notag  \\
Z_{+}(r_{*}) &=& e^{i\omega r_{*}}\sum_{j=0}^{\infty}\bar{\alpha}_{j}r^{-j},
\ena
where $Z_{-}~(Z_{+})$ represents the purely outgoing~(incoming) waves respectively.  The coefficient $\bar{\alpha}_{j}$ is the complex conjugate of $\alpha_{j}$, they can be found by the recursive relation~(e.g. in Ref.~\cite{Chandrasekhar:1975zza}) using conventional Frobenius method.  The generic solution in the outer region is then given by
\be
Z_{\rm out} = A(\omega)Z_{-} + B(\omega)Z_{+},  \label{sout}
\ee
where the ratio $\displaystyle{B(\omega)/A(\omega)}=1, 0$ for normal and quasinormal modes respectively.

\subsection{QNMs with small Im $\omega$}

A method although laborious yet effective in finding the QNMs with small imaginary parts is to scan for solution with $B(\omega)/A(\omega) = 0$ by varying $\omega$.  

\begin{figure}[h]
	\centering
	\includegraphics[width=0.5\textwidth]{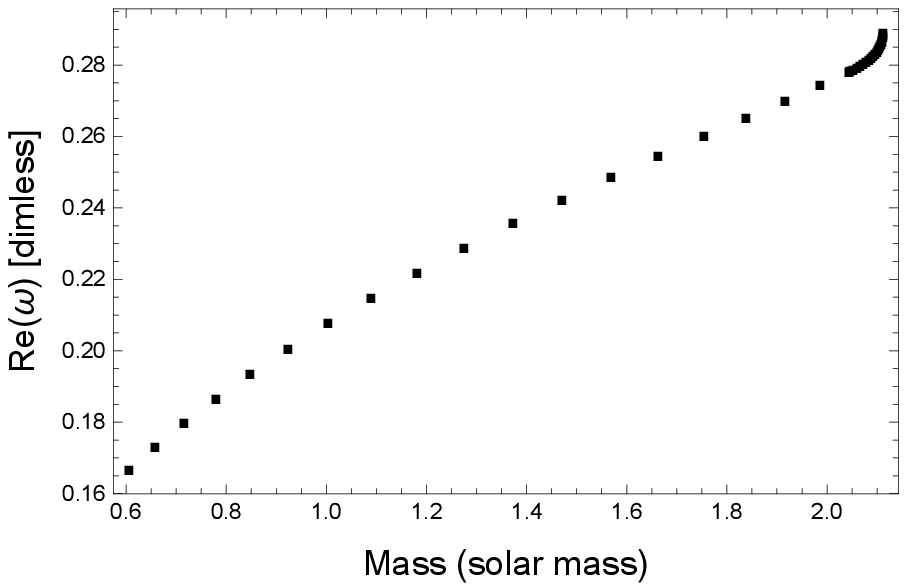}\\
      \includegraphics[width=0.5\textwidth]{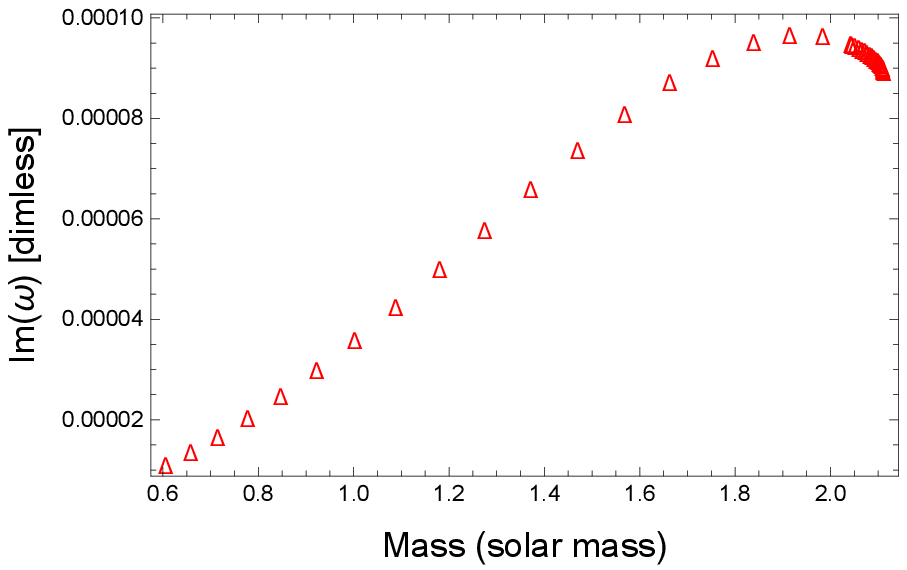}
	\caption{QNMs with small Im $\omega$ vs. mass of the MQ star. }
	\label{QNMfig1}
\end{figure}

The results are shown in Fig.~\ref{QNMfig1}.  These modes have less than $10^{-6}$ of the incoming/outgoing waves ratio at far distance $r=55 ~{\rm Re}~\omega$. As the MQ-star mass increases, the real and imaginary part of quasinormal frequencies increase with the MQ-star mass. Interestingly, polar QNMs of neutron star in massive scalar-tensor gravity also share a similar trend \cite{Blazquez-Salcedo:2021exm}.

\subsection{WKB}

\begin{figure}[h]
	\centering
	\includegraphics[width=0.5\textwidth]{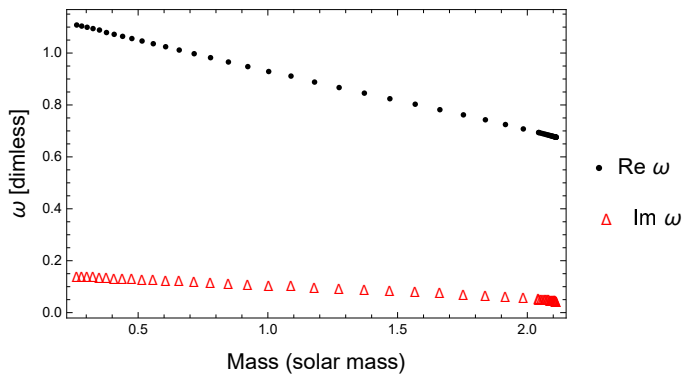}
	\caption{QNMs vs. mass of the MQ star. }
	\label{QNMfig2}
\end{figure}

For the WKB method, the approximate wave solution is given by replacement $Z_{\mp}\to Z_{\mp}^{\rm WKB}$ where
\be
Z_{\mp}^{\rm WKB}(r) = Q(r_{*})^{1/2}\exp\Big[ \mp i\int^{r_{*}}Q(y)dy \Big],
\ee
and  
\be
Q(r_{*})=(\omega^{2}-V(r))^{1/2} \notag
\ee
in Eq.~(\ref{sout}) respectively. The ratio of the incoming and outgoing waves is thus~\cite{Kokkotas:1992xak}
\bea
\frac{B(\omega)}{A(\omega)}&=& e^{-2iQ(R)}\frac{Q(R)-i\displaystyle{\Big[ \frac{Z'_{\rm in}(R)}{Z_{\rm in}(R)}+\frac{Q'(R)}{2Q(R)}\Big]}}{Q(R)+i\displaystyle{\Big[ \frac{Z'_{\rm in}(R)}{Z_{\rm in}(R)}+\frac{Q'(R)}{2Q(R)}\Big]}}, \label{wrat}
\ena
where the prime indicates derivative with respect to $r_{*}$. The QNMs can be determined by numerical evaluation of the roots of $B(\omega)/A(\omega) = 0$, by first substituting a trial value of $\omega$ and solve for the inner solution $Z_{\rm in}(r)$ within the star.  The resulting roots for $\omega$ is then fed back to the equation of motion to find a new inner solution and repeat the process until $\omega$ converges to a single value. Fig.~\ref{QNMfig2} shows QNMs of multiquark star with large ${\rm Im~\omega}$ determined by the WKB method, the physical value scales with $\sqrt{\epsilon_{s}}$. In contrast to Fig.~\ref{QNMfig1}, the real and imaginary part of quasinormal frequencies decrease monotonically with the MQ-star mass. Similar behavior is found for the axial perturbation of neutron star in $R^2$ gravity~\cite{Blazquez-Salcedo:2018qyy}.

\begin{figure}[h]
	\centering
	\includegraphics[width=0.5\textwidth]{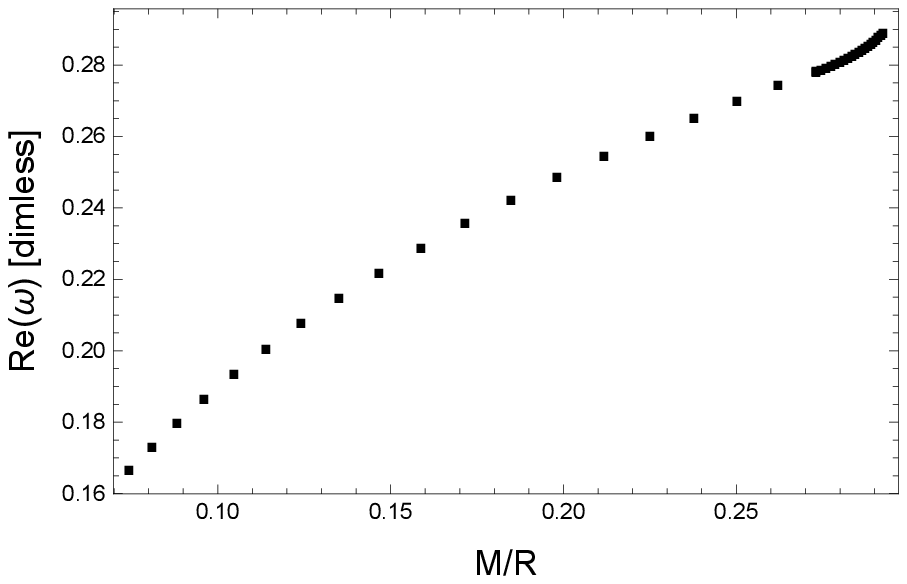}\\
      \includegraphics[width=0.5\textwidth]{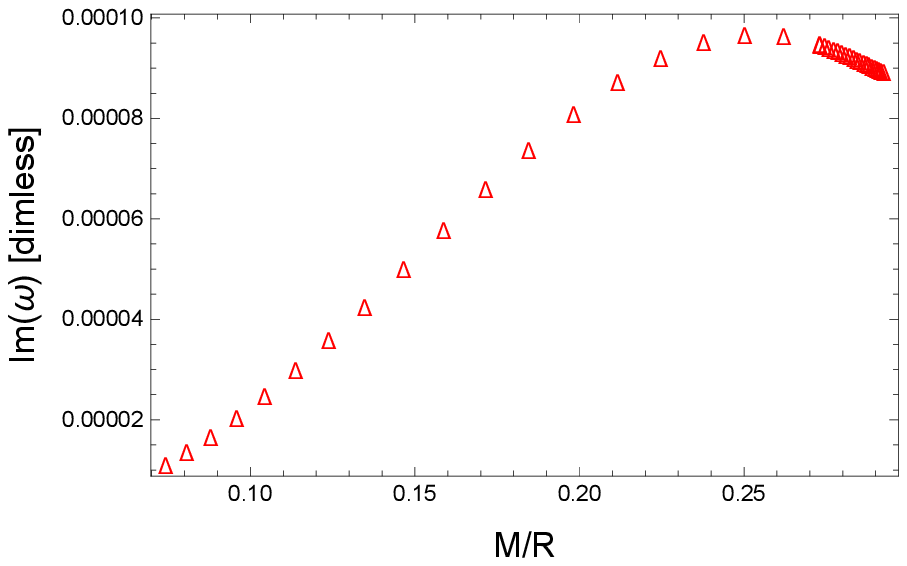}\\
	\includegraphics[width=0.5\textwidth]{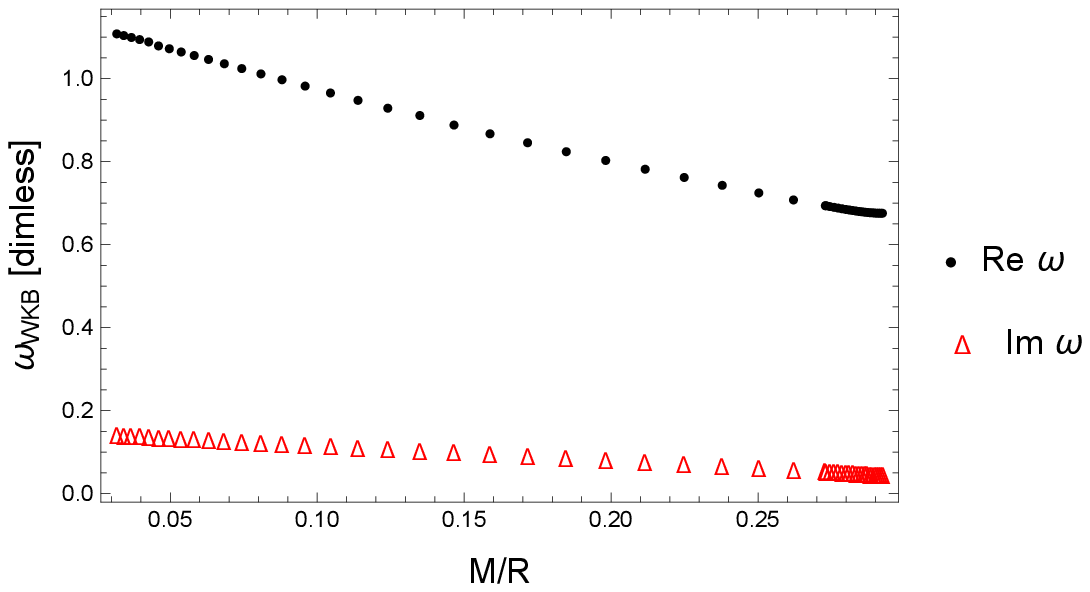}
	\caption{QNMs with small Im $\omega$ and large (WKB) Im $\omega$ vs. compactness $M/R$ of the MQ star. }
	\label{QCfig}
\end{figure}

In order to understand more on the physical origin of the two kinds of modes, we plot $\omega$ versus compactness $M/R$ in Fig.~\ref{QCfig}. The small-${\rm Im}~\omega$ modes have ${\rm Re}~\omega$ increasing with compactness while ${\rm Im}~\omega$ increases with $C$ until a mximum around $C\simeq 0.25-0.26$ then it decreases. In contrast, WKB modes~(with large ${\rm Im}~\omega$) have $\omega$ as a decreasing function of $C$ for both real and imaginary parts. ${\rm Re}~\omega$ and ${\rm Im}~\omega$ can be converted to SI units by conversion factors $f_{\rm con}$ and $t_{\rm con}$ respectively.

In addition, the small imaginary modes (the upper two plots in Fig.~\ref{QCfig}) appear to follow the universal relations of QNMs of neutron star reported in \cite{Tsui:2004qd} in a small compactness region i.e. $C < 0.25$. However, our results on the WKB modes do not share the universality found in \cite{Tsui:2004qd}. This discrepancy may allow us to distinguish the gravitational radiation from the MQ star.

\begin{figure}[h]
	\centering
	\includegraphics[width=0.5\textwidth]{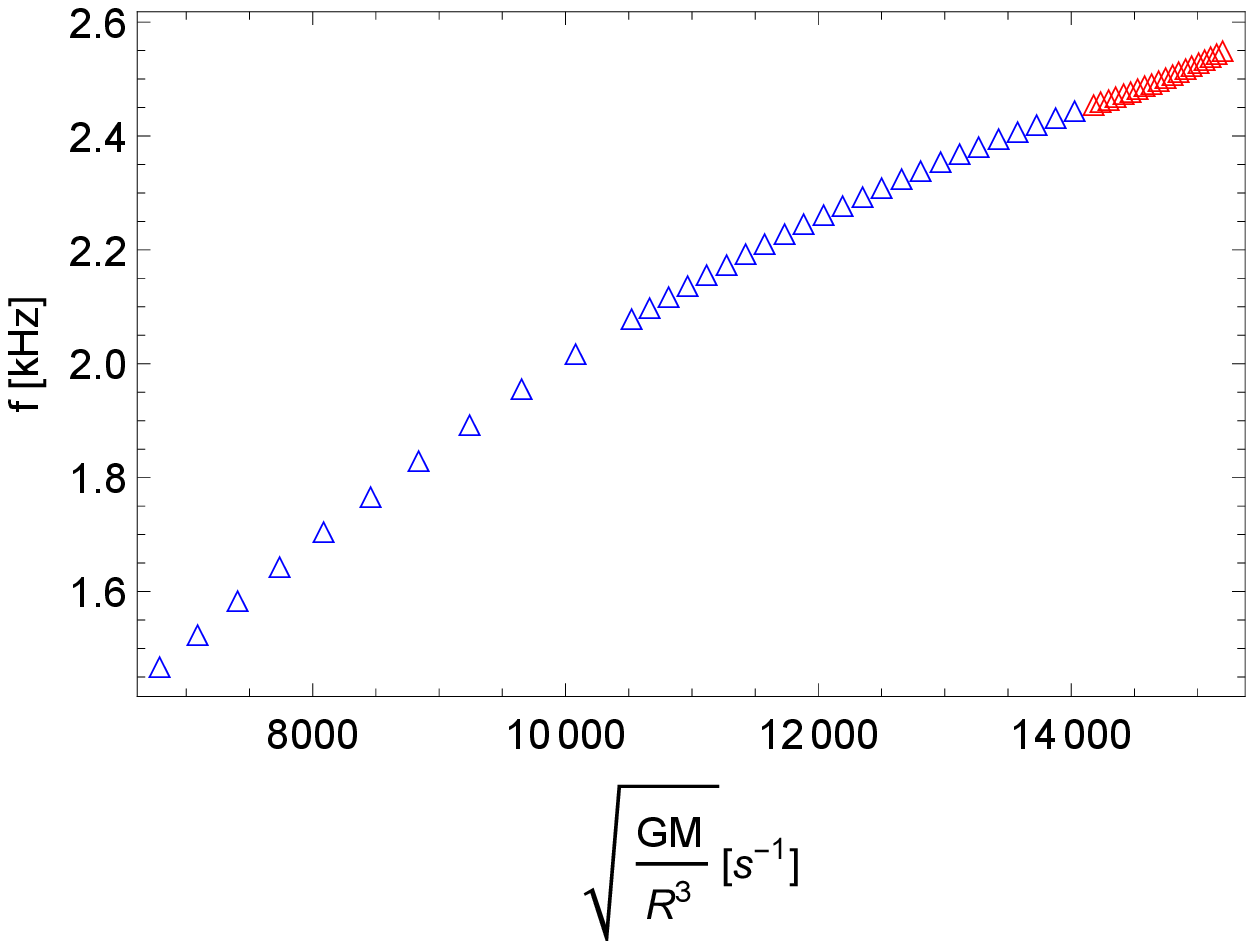}\\
      \includegraphics[width=0.5\textwidth]{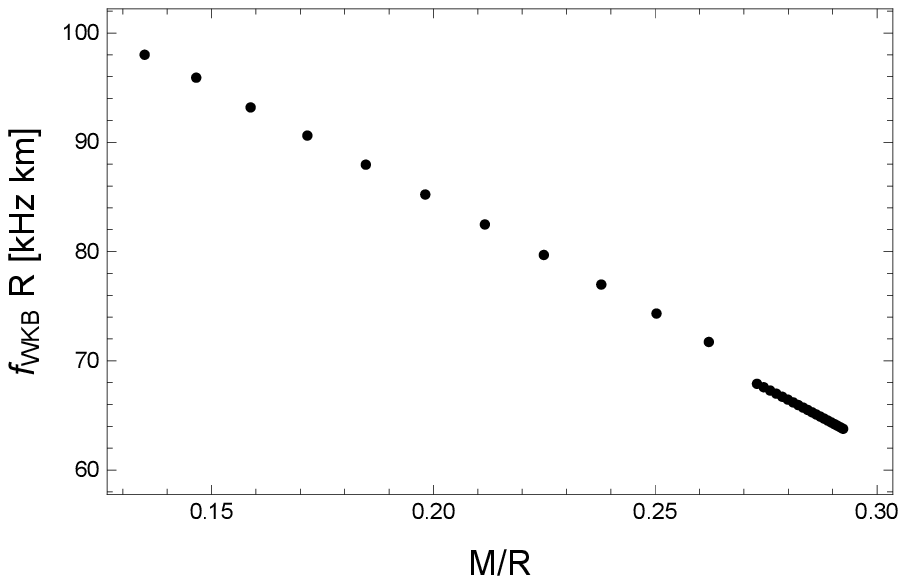}\\
	\caption{\raggedright{Frequencies $f = {\rm Re}~\omega/2\pi$ of $f-$modes vs. $\sqrt{G\bar{\rho}}$~(upper) and $fR$ of WKB $w-$modes vs. $C$~(lower). The red/dense dots on the high $C$ values show MQ star with high-density ``mqh" core.}}
	\label{modefig}
\end{figure}

\begin{figure}[h]
	\centering
	\includegraphics[width=0.5\textwidth]{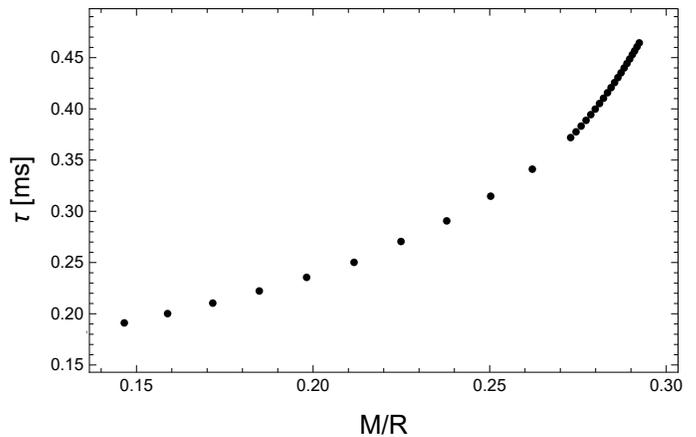}\\
     	\caption{\raggedright{Damping time $\tau$ of WKB $w-$modes vs. $C$. The dense dots on the high $C$ values show MQ star with high-density ``mqh" core.}}
	\label{modefig1}
\end{figure}

From the range of numerical values of the QNMs for both kinds and the approximate linear behaviour of the frequencies shown in Fig.~\ref{modefig}~($\bar{\rho}=M/R^{3}$, average density of the star), we can conclude that the small-${\rm Im}~\omega$ modes are the $f-$modes, and the WKB QNMs are the curvature $w-$modes according to the criteria in Ref.~\cite{Kokkotas:1999bd,Andersson:1997rn}. Moreover, the damping time of the WKB modes is an increasing function of compactness as shown in Fig.~\ref{modefig1}, consistent with the property of the curvature $w-$modes.  The spacetime metric perturbations $H_{1}, K$ of these modes clearly are dominant as shown in Appendix~\ref{app}.

\section{Conclusions and Discussions}\label{sec-con}

The oscillatory modes of holographic MQ star/core of massive stars in the SS model with the energy density scale $\epsilon_{s}=26$ GeV fm$^{-3}$ have been calculated.  We obtain normal and quasinormal modes of MQ star by using direct scan and WKB methods. We find that the stars with higher mass and compactness oscillate with higher frequencies. By direct scanning of the solutions satisfying boundary condition of the quasinormal modes, i.e., zero incoming gravitational waves from infinity in the outer region of star, QNMs with very small ${\rm Im}~\omega$ are found with frequencies in the order of $1.5-2.6$ kHz, and damping times $0.19-1.7$ secs for MQ star with mass $M=(0.6-2.1) M_{\odot}$~(note that the MQ star with mass smaller than $1.4 M_{\odot}$ is most likely hypothetical, however we choose to present them for comparison to the typical NS with other nuclear EoS). By using WKB method, QNMs with larger ${\rm Im}~\omega$ are found with $f=5.98-9.81$ kHz, damping times $0.13-0.46$ ms for $M\simeq (0.3-2.1) M_{\odot}$. These are the $f-$modes and curvature $w-$modes of MQ star respectively. Both modes can be fit with approximate empirical linear relations found in Ref.~\cite{Andersson:1997rn} as shown in Fig.~\ref{modefig}. For MQ $f-$modes, since the EoS is not a single power law, the approximate linear relation can fit well only around the high mass region of the star in Fig.~\ref{modefig}.  For $w-$modes, the linear relation fit is quite excellent even though there appears to be a transition from one fitting linear relation to another when the MQ star changes from the star with high density ``mqh" core to pure ``mql'' star.  

Massive neutron star around and above two solar masses could have multiquark core whence further gravitational collapse would generate fluid and spacetime excitations in the $f, p, g, r$ and $w-$modes.  The GW from these excitations could be detected after such collapse e.g. in aftermath of supernovae explosion and neutron stars collision.  It is thus possible that these QNMs would be generated by such extreme events and subsequently detected at LIGO/Virgo and future gravitational waves detection facilities. The sensitivities of these GW events are estimated in Appendix~\ref{app1}.    

\begin{acknowledgments}

SP~(first author) was supported by Grant No. RGNS 64-217 from Office of the Permanent Secretary,
Ministry of Higher Education, Science, Research and Innovation  (OPS MHESI),
Thailand Science Research and Innovation (TSRI) and Silpakorn University. PB is grateful to APCTP for warm hospitality during the visit where part of this work has been done. 

\end{acknowledgments}

\appendix

\section{Perturbation profiles of WKB $H_{1}(r), K(r), W(r), X(r)$ inside the MQ star}   \label{app}

\begin{figure}[h]
	\centering
	\includegraphics[width=0.5\textwidth]{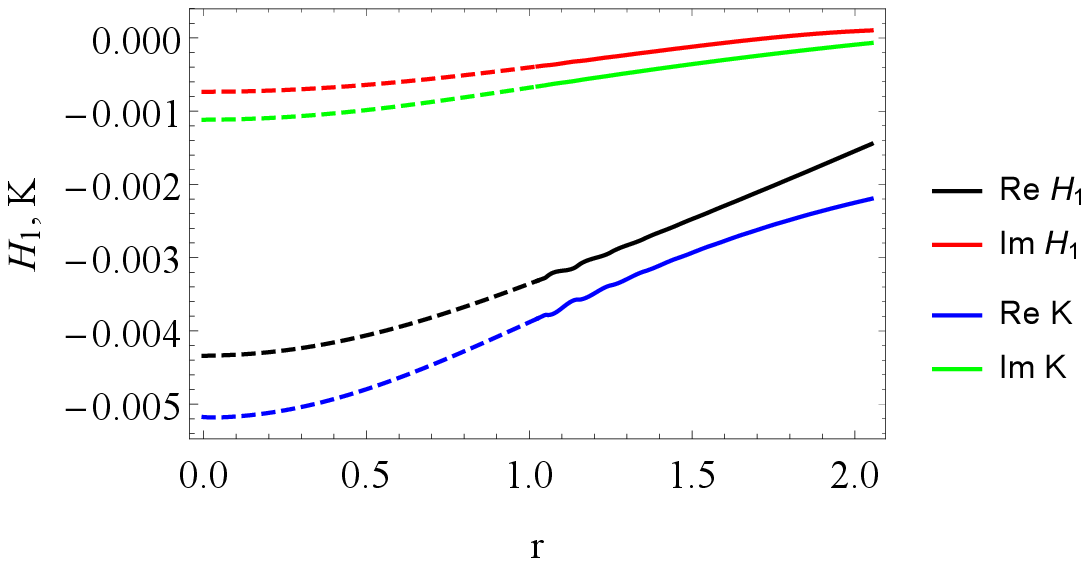}\\
      \includegraphics[width=0.5\textwidth]{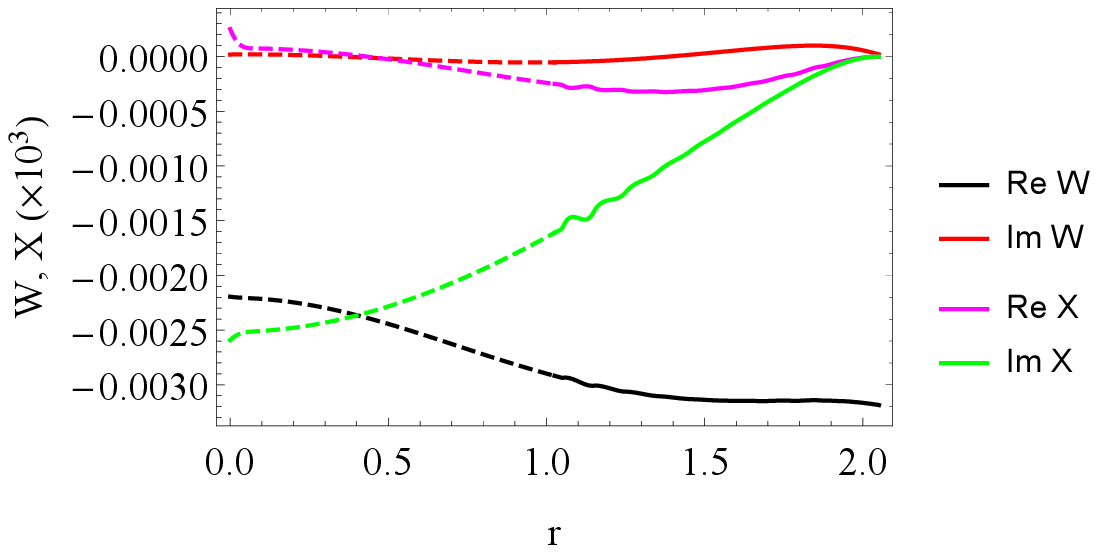}\\
	\caption{\raggedright{Perturbations inside MQ star for WKB mode at $M=2.04~M_{\odot}$, note that the value of $X$ is multiplied by $10^3$.}}
	\label{profileWKBfig}
\end{figure}

Perturbations inside the MQ star for the WKB modes~(spacetime curvature $w-$modes) at $M=2.04~M_{\odot}$~(where ``mqh" disappears and only ``mql" exists) MQ star are shown in Fig.~\ref{profileWKBfig}. The spacetime metric perturbations $H_{1}, K$ are clearly dominant for this mode. 

\section{Sensitivity of GW signals from $f-$mode and $w-$mode oscillations of MQ star}   \label{app1}

In order to estimate the sensitivities of GW from oscillating MQ star at the detection facilities, we adapt the approximation formula for GW amplitude $h_{\rm eff}$ by Andersson and Kokkotas~\cite{Andersson:1996pn,Andersson:1997rn},
\be
h_{\rm eff} \sim 2.2\times 10^{-21}\Big(\frac{E}{10^{-6}M_{\odot}c^{2}}\Big)^{1/2}\Big(\frac{2{\rm kHz}}{f}\Big)^{1/2}\Big(\frac{50 {\rm kpc}}{r}\Big),
\ee
and
\be
h_{\rm eff} \sim 9.7\times 10^{-22}\Big(\frac{E}{10^{-6}M_{\odot}c^{2}}\Big)^{1/2}\Big(\frac{10{\rm kHz}}{f}\Big)^{1/2}\Big(\frac{50 {\rm kpc}}{r}\Big)
\ee
for the $f-$mode and $w-$mode respectively. $E$ is the available energy in GW form and $r$ is the distance to the source.  

The sensitivity at a detector is then given by strain$/\sqrt{\rm Hz} \simeq h_{\rm eff}/10\sqrt{\rm Hz}$ for signal bandwidth 100 Hz. For $M=M_{\rm max}=2.11 M_{\odot}$, the lowest $f-$mode~($w-$mode) $f=2.55~(5.98)$ kHz, the required minimum sensitivity for detection is $1.54~(1.0)\times 10^{-24}/\sqrt{\rm Hz}$ for $E=10\times 10^{-6}M_{\odot}c^{2}, r=20$ Mpc respectively. Large uncertainties come from the amount of available energy in each mode of GW since $h_{\rm eff}\sim \sqrt{E}$. With our estimate of $E=10\times 10^{-6}M_{\odot}c^{2}$and by comparison to Table~\ref{tabII}, the GW signals in these modes are still beyond the discovery sensitivity of present detection facilities, LIGO/Virgo, for nearby source at $20$ Mpc~(half the distance of GW170817). However, the future detection facilities such as the Einstein Telescope and Cosmic Explorer have the potential to discover GW signals from these modes for events with $E\sim 10\times 10^{-6}M_{\odot}c^{2}$ at $20$ Mpc.
\begin{table}[ht]
	\centering	
	\begin{tabular}{ |c|c|c|c| }
		
		\hline
             &&&\\
		minimum&Advanced &Einstein &Cosmic \\
		
		sensitivity($1/\sqrt{\rm Hz}$)&LIGO/Virgo&Telescope&Explorer \\
		
		(at)&&&\\
		
		\hline
		&&&\\
		$2.55$ kHz&$\gtrsim 1\times 10^{-23}$&$\gtrsim 1\times 10^{-24}$&$1\times 10^{-24}$\\
		&&&   \\

		\hline
		&&&\\
		$5.98$ kHz&$\gtrsim 3\times 10^{-23}$&$\gtrsim 3\times 10^{-24}$&$2\times 10^{-24}$\\
		&&& \\

		\hline
		
	\end{tabular}
	
	\caption{Minimum sensitivity at the lowest $f-$mode and $w-$mode MQ star frequencies at present and future detection facilities estimated from Fig. 1 of Ref.~\cite{Martynov:2019gvu}}
	
	\label{tabII}
	
\end{table}

\end{document}